\documentclass[pdftex,twocolumn,epjc3]{svjour3}          

\RequirePackage[T1]{fontenc}

\smartqed  

\RequirePackage{graphicx}
\RequirePackage{mathptmx}      
\RequirePackage{flushend}
\RequirePackage[numbers,sort&compress]{natbib}
\RequirePackage[colorlinks,citecolor=blue,urlcolor=blue,linkcolor=blue]{hyperref}

\journalname{Eur. Phys. J. C}

\usepackage{slashed}
\usepackage{amssymb}

\newcommand{\Dslash}{\slashed{D}}
\newcommand{\Dslashbak}{\overleftarrow{\Dslash}}

\begin{document}

\title{Finite Temperature Landau Gauge Lattice Quark Propagator}


\author{Orlando Oliveira\thanksref{e1}
        \and
        Paulo J. Silva\thanksref{e2} 
}

\thankstext{e1}{e-mail: orlando@uc.pt}
\thankstext{e2}{e-mail: psilva@uc.pt}

\institute{CFisUC, Departamento de F\'{\i}sica, Universidade de Coimbra, 3004-516 Coimbra, Portugal}

\date{Received: date / Accepted: date}

\maketitle

\begin{abstract}
The quark propagator at finite temperature is investigated using quenched gauge configurations. The
propagator form factors are investigated for temperatures above and below the gluon deconfinement temperature $T_c$
and for the various Matsubara frequencies. Significant differences between the functional behaviour below and above
$T_c$ are observed both for the quark wave function and the running quark mass. 
The results for the running quark mass indicate a strong link between gluon dynamics, the mechanism for chiral symmetry 
breaking and the deconfinement mechanism. For temperatures above $T_c$ and for low momenta,
our results support also a description of quarks as free quasi-particles.
\end{abstract}

\section{Introduction and Motivation}

The study of strong interactions including temperature and density effects has been driven both by the
tentative to understand the QCD dynamics and by a strong experimental program that involves various facilities.
From the theory side, this research includes the computation of two-point correlation functions that provide
information on spectra, transport properties and other fundamental properties such as the confinement
mechanism. 
Herein, we will not consider the dependence on the density of hadronic matter.

At low temperatures quarks and gluons are confined particles and appear only as constituents of mesons and
baryons. At sufficiently high temperatures or densities quark and gluons are expected to become deconfined and
behave as free quasiparticles in a new state of matter, the strong coupled quark gluon plasma. 
At extreme temperatures, due to asymptotic freedom, quarks and gluons can be considered as free particles. 

This view of the quark and gluon dynamics for temperatures above the critical temperature $T_c$, where
the deconfined phase transition takes place, is motivated by asymptotic freedom and lies at the heart of
the quasiparticle description used to investigate the thermodynamical properties of hadronic matter for $T > T_c$. 
For temperatures above $T_c$ the perturbative approach of the hard thermal loop (HTL) expansion, 
see e.g.~\cite{LeBellac:1991cq,Kapusta:2006pm} and~\cite{Su:2015esa} for a recent review, seems to be the
framework to understand the dynamics of QCD. Furthermore, some of the HTL predictions have been
confirmed by first principles non-perturbative lattice QCD simulations; an example being the behaviour of thermal masses with 
the temperature of the heat bath.

The standard approach to identify the confined and the deconfined phases relies on the Polyakov loop $L$ which is related to
the free energy of quarks $F_T$.
At $T < T_c$, the renormalised $L$ is small, $F_T$ is large, and quarks are confined particles. 
For temperatures above $T_c$, $L$ approaches unit, $F_T \approx 0$, and quarks become free quasiparticles. 
The curve $L(T)$ shows that the nature of the confined-deconfined transition is a cross-over for full QCD and 
first order for the pure Yang-Mills (quenched QCD)~\cite{McLerran:1980pk,McLerran:1981pb}. 
If for full QCD the critical temperature is  $T_c \sim 150-160$ MeV, for its quenched version one gets $T_c \sim 270$ MeV. 
In what concerns gluon dynamics, for the quenched version of the theory, the consideration of other quantities besides 
the Polyakov loop reproduce both the nature of the transition and the quoted value for the critical temperature, see e.g.~\cite{Silva:2016onh}.

One of the interests to study the quark and gluon two-point correlation functions with temperature comes from
the connection between the propagators and confinement that can be linked, for example, by computing the 
corresponding spectral functions. If above $T_c$ quarks and gluon can be viewed as quasiparticles, their propagators 
should behave differently from the corresponding functions at temperatures below $T_c$.

The Landau gauge gluon propagator has been studied using non-pertur\-ba\-ti\-ve methods for a wide range of temperatures both for full QCD
and for the pure gauge theory~\cite{Huber:2018ned,Aouane:2011fv,Silva:2013maa,Aouane:2012bk,Silva:2016msq,Silva:2016onh}. 
These studies show that in the quenched theory the gluon propagator is sensitive to the breaking of center symmetry, 
that electric and magnetic mass scales are generated dynamically and the gluon thermal mass associated with the electric propagator 
scales with the temperature according to the prediction of hard thermal loop approach to QCD,  namely $m_D \propto T$ above the 
critical temperature. 
Furthermore, the calculations performed using the continuum formulation of QCD
or lattice QCD simulations for the pure gauge sector are in good agreement; see~\cite{Maas:2011se} and references therein.
Moreover, the non-perturbative approaches suggest that HTL QCD is a good framework to describe the dynamics of QCD for 
$T \sim  T_c$ and above. From these studies one can claim to have now a good picture for the gluon
dynamics. 

The Landau gauge quark propagator at finite temperature was also studied within the continuum non-perturbative approaches to QCD and
using first principles lattice simulations. 

The quark gap equation in the Landau gauge was solved relying on gluon propagators obtained
from lattice simulations. The lattice data was fitted to a functional form that reproduces the perturbative tail at sufficiently high 
momenta~\cite{Ikeda:2001vc,Mueller:2010ah,Fischer:2010fx,Contant:2017gtz}. 
In order to solve the Dyson-Schwin\-ger equation for the quarks, the quark-gluon vertex was parameterised keeping keeping only its tree
level tensor structure, i.e. assuming $\Gamma_\mu \propto \gamma_\mu$ for the vertex, and, once more, ensuring the right perturbative tail 
at high momenta. 
The continuum studies have been focused on dynamical mass generation, through the computation of either
the quark condensate~\cite{Fischer:2009wc}, the running quark mass, 
the spectral function~\cite{Qin:2010pc} at the chiral limit and order parameters for the deconfined phase transition. 
The form factors appearing on the quark propagator as function of the Matsubara frequencies, momentum and temperature have been 
only briefly reported but they show quite different behaviours above and below the critical temperature. How the various functions 
that appear on the quark propagator depend on the parameterisations introduced to solve the quark gap equation is not known 
and, therefore, the results obtained care for further study or confirmation from independent calculations. 
The interest on the spectral functions at the chiral limit comes from knowing that
at high temperature the quark propagator has two sets of poles, corresponding to an usual mass term and a collective
plasmino mode, i.e. two different types of dispersion relations can be associated with this propagator~\cite{Weldon:1999th}. 
Some authors also speculate on the presence of a third ultrasoft fermion mode that should appear on the quark 
propagator~\cite{Qin:2010pc,Nakkagawa:2011ci,Nakkagawa:2012ip}.
Evidence for a third  mode for $T > T_c$ was also observed on Yukawa like models~\cite{Harada:2008vk}, on
QED like models where it becomes an ultrasoft mode~\cite{Satow:2010ia,Hidaka:2011rz} and in 
low energy effective models of QCD~\cite{Kitazawa:2005mp,Kitazawa:2006zi,Kitazawa:2006vh,Kitazawa:2007ep} where, again, 
the third mode is ultrasoft. 

In what concerns continuum methods, the quark spectral function together with the quark self energy was
computed within the framework of functional renormalization group~\cite{Wang:2018osm}. The outcome reveals a
spectral function with multiple peaks, whose details and number of maxima depend on the truncation used in the calculation.

The lattice quark propagator at finite temperature, in the Landau gauge, was computed for $\mathcal{O}(a)$-improved Wilson fermions
using quen\-ched gauge configurations by a number of 
authors~\cite{Hamada:2006ra,Karsch:2007wc,Karsch:2007bf,Hamada:2008zz,Karsch:2009tp,Kitazawa:2009uw,Hamada:2010zz,Kaczmarek:2012mb}.
The lattice studies have been focused either on the mass function and on the same spectral function as investigated by the continuum methods.
For the quantities studied, at the qualitative level, lattice and continuum results are in good agreement. On the lattice 
the curve $M/T$ as a function of the temperature, where $M$ is the quark mass, has a discontinuity at $T_c$ but its behaviour for $T > T_c$
has not been resolved~\cite{Hamada:2010zz}. For Wilson fermions, the authors~\cite{Hamada:2006ra} found a linear grow of
$M/T$ above the transition temperature. However, given that there is some ambiguity on the definition of the quark mass and the poor agreement
between the values of the Wilson fermion mass and the $\mathcal{O}(a)$-improved Wilson fermion mass, this result demands for confirmation.
Furthermore, in what concerns the computation of the quark mass, a plateaux associated to an effective quark mass was not
always identified~\cite{Hamada:2006ra,Hamada:2008zz} suggesting that instead one should consider a running quark mass.
The lattice spectral function was computed and investigated assuming a multiple pole 
\textit{ans\"atze}~\cite{Karsch:2007wc,Karsch:2007bf,Karsch:2009tp,Kitazawa:2009uw,Kaczmarek:2012mb} that favours
a pole structure associated with a thermal and a plasmino mode as predicted in~\cite{Weldon:1999th}. The lattice calculations for
the propagator used typically spatial physical volumes under ($\sim$2 fm)$^3$ with two simulations performed on physical volumes
($\sim$3.5 fm)$^3$~\cite{Karsch:2009tp,Kaczmarek:2012mb} for $T/T_c = 0.55$ and 1.5 at the chiral limit\footnote{The authors
call the reader's attention that in these simulations they used $T_c = 300$ MeV for the transition temperature (quenched theory) 
and the current used value is $T_c = 270$ MeV. A rescaling of the ratios just quoted results in $T/T_c =  0.61$ and 1.72.}.

In the current work we compute the lattice Landau gauge quark propagator at finite temperature with non-pertur\-ba\-ti\-ve\-ly 
$\mathcal{O}(a)$-improved Wilson fermions with gauge configurations generated with the Wilson action for the pure gauge 
theory. These configurations were used to investigate the lattice Landau gauge gluon propagator in~\cite{Silva:2013maa}. 
As in this work, we take $T_c = 270$ MeV and consider temperatures above and below the critical temperature to study 
how different the propagator is in the confined and deconfined phase. 
Our main focus is on the calculation of the three form
factors that determine the quark propagator at finite temperature and the running quark mass
as a function of momentum and for the various Matsubara frequencies. By providing the form factors as a function of $T$,
we aim to understand the difference between the two phases and provide information than can be useful also for the
continuum approaches to QCD. No attempt is made to investigate any of the spectral functions that can be associated with
the quark propagator.
 
The computations of the quark propagator reported below use lattices whose spatial physical volume is ($\sim$ 6.5 fm)$^3$ and
various lattice spacings around $a \sim 0.1$ fm. 
The lattice investigations of the Landau gauge gluon and ghost propagators
at zero temperature~\cite{Oliveira:2012eh,Duarte:2016iko,Dudal:2018cli}, suggest that finite volume effects
are under control in the sense that they are small and below the statistical precision of the simulations.
We take as definition of the temperature the inverse length on the time direction and typically consider ratios
of the spatial and time direction $L_s / L_t = 8$. Due to the anti-periodic boundary conditions along the time direction
$p_4 = \pi \, T \, ( 2 \, n_t + 1 )$ where $n_t = 0, \, 1, \, \dots$
and the spatial momenta are
$p_i  = 2 \, \pi \, T \, (L_t / L_s) \,  n_i \approx 0.785 \, T  \,  n_i $ for $n_i = 0, \, 1, \dots$
Furthermore, in order to experience the quark propagator at the chiral limit, we report
on simulations using two values of the bare quark mass, namely $m_0 \approx 10$ MeV and 50 MeV.

Our results show clearly that the nature of the quark propagators form factors changes for temperature above
$T_c$, compared with temperatures in the confined phase. This is reported with detail for both the quark wave function
and the running quark mass. 
In particular, for the running quark mass we find that it is highly suppressed above $T_c$,
with typical infrared values being about half of the corresponding values for the temperatures below $T_c$.
Our computation if performed with quenched configurations and this results indicates that the gluon dynamics plays
an  important role on the mechanism of chiral symmetry breaking. 
A similar change on the functional form of the finite temperature gluon propagator form factors 
for temperatures above and below $T_c$ was observed also in lattice simulations~\cite{Aouane:2011fv,Silva:2013maa}.
Indeed, these simulations show that the relative importance of its electric and magnetic components are inverted for $T \gtrsim T_c$, 
relative to confined phase, i.e. for small temperatures the electric form factor is larger than the magnetic one, while above $T_c$
the dominant form factor is associated with the gluon magnetic component.

The paper is organised as follows. In Sec.~\ref{Sec:Def} we set the notation and definitions used through out the current
work and how the various form factors are measured. Further, we detailed the setup of the simulations analysed herein.
In Sec.~\ref{Sec:Prop} we report our results for the various form factors prior to the estimation of the lattice artefacts
for the two quark masses considered in the simulations. The study of the lattice artefacts and the definition of the running
quark mass are detailed in Sec.~\ref{Sec:LattArt}.

\section{The Quark Propagator and Lattice Setup \label{Sec:Def}}

\begin{table*}
\begin{center}
\begin{tabular}{cccllllccc}
\hline
T  &  $\beta$  &  $L_s^3 \times L_t$  & $\kappa$ & $\kappa_c$ & $a$ & $1/a$ & $m_{bare}$ & $c_{sw}$  \\
\hline
243   & 6.0000     &  $64^3\times8$   & 0.1350 &  0.13520 &0.1016 & 1.9426 & 10   & 1.769\\
      &            &                  & 0.1342 &     &  & & 53   &      \\
\hline
260   & 6.0347     &  $68^3\times8$   & 0.1351 & 0.13530 & 0.09502 & 2.0767 & 11   & 1.734\\
      &            &                  & 0.1344 &  & & & 51   &      \\
\hline
275   & 6.0684     &  $72^3\times8$   & 0.1352 & 0.13540 & 0.08974 & 2.1989 & 12   & 1.704\\
      &            &                  & 0.1345 & & & & 54   &      \\
\hline
290  & 6.1009     & $76^3\times8$    & 0.1347  & 0.13550  & 0.08502  & 2.3211 &  51  & 1.678 \\
\hline
305  & 6.1326     & $80^3\times8$    &  0.1354   &  0.13559 &  0.08077  & 2.4432  & 13  & 1.655 \\      
        &                 &                             & 0.1348  &                &                &            &  53  &   \\
\hline
324   & 6.0000     & $64^3\times6$   & 0.1342  & 0.13520 & 0.1016 & 1.9426 &  53   &   1.769 \\      
\hline      
\end{tabular}
\end{center}
\caption{Lattice setup for the computation of the quark propagator in the quenched approximation. The
coefficient $c_{sw}$ associated with the non-perturbative improvement of the simulations and $\kappa_c$
are taken from~\cite{Luscher:1996sc}. The bare quark mass $m_{bare}$ is computed using Eq. (\ref{Eq:mbare}).
$T$ and $m_{bare}$ are given in MeV, the lattice spacing $a$ is given in fm and the inverse of
the lattice spacing $1/a$ in GeV.} \label{Tab:LatSet}
\end{table*}

On the continuum, the quark propagator is diagonal in color space. At finite temperature, the presence of a thermal
bath breaks rotational invariance, and the inverse of the space-spin quark propagator in momentum space reads
\begin{equation}
   S^{-1}(p_4 , \vec{p} ) 
    =  i \gamma_4 \, p_4 ~ \omega (p_4, \vec{p}) + i \vec{\gamma} \cdot \vec{p} ~ Z(p_4 , \vec{p} ) + \sigma (p_4 , \vec{p} )  
    \label{Eq:InvContQuarkProp}
\end{equation}   
or, equivalently,
\begin{equation}
   S(p_4 , \vec{p} )  = \frac{ - i \gamma_4 \, p_4 ~ \omega (p_4, \vec{p}) - i \vec{\gamma} \cdot \vec{p} ~ Z(p_4 , \vec{p} ) + \sigma (p_4 , \vec{p} ) }
                                        {   p^2_4 ~  \omega^2 (p_4, \vec{p})   + \left( \vec{p}\cdot\vec{p} \right) ~ Z^2(p_4 , \vec{p} )  + \sigma^2 (p_4 , \vec{p} ) }.
                                        \label{Eq:ContQuarkProp}
\end{equation}
Our analysis of the lattice propagator will assume that we are close to the continuum and, therefore, expressions
(\ref{Eq:InvContQuarkProp}) and (\ref{Eq:ContQuarkProp}) can be applied. The form factors
$\omega$, $Z$ and $\sigma$ can be accessed by computing traces of the propagator times gamma matrices.

Our calculation of the lattice quark propagator relies on the use of the non-perturbative $\mathcal{O}(a)$ improved
clover action~\cite{Sheikholeslami:1985ij,Luscher:1996sc}
with tree-level O(a)-improved sources~\cite{Heatlie:1990kg,Skullerud:2000un,Skullerud:2001aw}.
The lattice quark propagator in real space reads
\begin{equation}
    S(x,y) = \bigg( 1 + 2 \, b_q \, a \, m \bigg) ~
    \left\langle ~ L(x) \, M^{-1}_{SW} (x,y) \, R(y)  ~ \right\rangle \ , 
    \label{EQ:LatProp}
\end{equation}
where $M^{-1}_{SW}$ stands for the inverse of the improved fermionic matrix, the rotated sources are given by
\begin{eqnarray}
   L(x) & = & \left[ 1 - c_q \, a \, \overrightarrow{\Dslash} (x)  \right]  \ ,\\
   R(x) & = & \left[ 1 + c_q \, a \,  \Dslashbak (x) \right] \ ,
\end{eqnarray}
the left and right derivatives are
\begin{eqnarray}
\overrightarrow{\Dslash} (x) \, \psi (x) \!\! & = & \!\!
         \sum_\mu \frac{\gamma_\mu}{2 a} \left[ U_\mu (x) \psi ( x + \hat{\mu} )  -  U^\dagger_\mu ( x - \hat{\mu} ) \psi (x - \hat{\mu}) \right]
          ,
         \nonumber \\
         & & \\
\overline \psi (x)  \Dslashbak (x)  \!\! & = & \!\! \sum_\mu \left[ \overline \psi ( x + \hat{\mu} ) U^\dagger_\mu (x) -
  \overline \psi ( x - \hat{\mu}) U_\mu ( x - \hat{\mu} ) \right] \frac{\gamma_\mu}{2 a} ,\nonumber \\
\end{eqnarray}
$U_\mu (x)$ stands for the gauge configuration, $\hat{\mu}$ is the unit vector associated with direction $\mu$ and $a$ is the lattice
spacing. For the coefficients associated with the improvement coming from the rotated sources we use their
tree level value $b_q = c_q = 1/4$. 

The momentum space propagator is obtained from Eq. (\ref{EQ:LatProp}) after a
Fourier transformation. Recall that on the lattice the fermionic boundary conditions are periodic in the spatial directions and 
anti-periodic in time and the available momenta $\vec{p}$ and $p_4$ assume the following discrete values 
\begin{equation}
   p_4 = \frac{  \pi}{L_t} \, \left( 2 \, n + 1 \right) \qquad\mbox{ and }\qquad
   p_i = \frac{ 2 \, \pi}{L_s} \,  n \ , 
\end{equation}
where $n = 0, \, 1, \, 2, \, \dots$

On the lattice the bare quark mass is given by
\begin{equation}
 a \, m = \frac{1}{2} \left( \frac{1}{\kappa} - \frac{1}{\kappa_c} \right) \ ,
 \label{Eq:mbare}
\end{equation}
where the critical hopping parameter $\kappa_c$ depends on $\beta$, i.e. on the lattice spacing,
and is defined as the value of $\kappa$ corresponding to a vanishing mass for the lightest pseudo-scalar meson. 
In the following we will use $m$ as an indicator of how close  the simulation is from the chiral limit. 
The $\kappa_c$ are taken from \cite{Luscher:1996ug}, interpolating their reported values  when necessary.

The traces of propagator give us the form factors
\begin{eqnarray}
  \frac{ \omega (p_4, \vec{p})}
         {   p^2_4 ~  \omega^2 (p_4, \vec{p})   + \left( \vec{p}\cdot\vec{p} \right) ~ Z^2(p_4 , \vec{p} )  + \sigma^2 (p_4 , \vec{p} ) } \  ,  
         \label{EQ:LatOmega} \\
  \frac{ Z(p_4 , \vec{p} )}
        {   p^2_4 ~  \omega^2 (p_4, \vec{p})   + \left( \vec{p}\cdot\vec{p} \right) ~ Z^2(p_4 , \vec{p} )  + \sigma^2 (p_4 , \vec{p} ) } \ , 
         \label{EQ:LatZ} \\
  \frac{\sigma (p_4 , \vec{p} ) }
 {   p^2_4 ~  \omega^2 (p_4, \vec{p})   + \left( \vec{p}\cdot\vec{p} \right) ~ Z^2(p_4 , \vec{p} )  + \sigma^2 (p_4 , \vec{p} ) },
  \label{EQ:LatM}
\end{eqnarray}
and by taking ratios of these functions one gets a continuum-like quark wave function and running quark mass defined by
\begin{eqnarray}
   Z_c(p_4, \vec{p})  =  \frac{Z(p_4, \vec{p})}{\omega (p_4, \vec{p})}  ,  \quad \mbox{ and } \quad
   M (p_4, \vec{p})  =  \frac{\sigma(p_4, \vec{p}) }{ \omega(p_4, \vec{p}) } ,
\end{eqnarray}   
respectively. Note that we use $\omega$ and not $Z$ to define the ratios $Z_c$ and $M$ because the simulation does not
allow the computation of $Z(p_4, \vec{p} = 0)$.

For the computation of the Landau gauge quark propagator we use a subset of the quenched gauge configurations, rotated to the Landau gauge,
generated for the work~\cite{Silva:2013maa}. We refer the read to this paper for the details on the generation of the Wilson action
pure gauge configurations, the rotation to the Landau gauge and on the scale setting for the conversion into physical
units\footnote{As discussed in~\cite{Boucaud:2017ksi,Duarte:2017wte} there is a small uncertainty in the definition of the lattice spacing
that can havw a small impact on the outcome specially for large statistical ensembles of configurations.
In the current work, we do not take into account this uncertainty.}. Furthermore, in order to minimise effects due to the breaking
of the rotational invariance, we only report functions for momenta that verifies the cuts considered in~\cite{Silva:2013maa}
and introduced in~\cite{Aouane:2011fv}; see also~\cite{Leinweber:1998uu}. In order to improve the signal to noise ratio
we perform always a $Z_3$ average of the various lattice quantities; for example when reporting a quantity
associated $F$ with momenta $(p_4, p, 0, 0)$ it refers to $( F(p_4, p, 0, 0) + (p_4, 0, p, 0) + (p_4, 0, 0, p))/3$.

The statistical errors reported for the various quantities were computed using  the bootstrap method
with a confidence level of 67.5\%. When reporting the `continuum'' functions $Z_c$ and $M$,
that are defined as ratios of traces of quark propagator multiplied by gamma matrices, an additional cut on the data
surviving the momenta cuts is introduced to have a clear picture on the various functions and only the data whose
relative error is below 50\% is kept.

The lattice setup of the simulations reported below are given in Tab. \ref{Tab:LatSet}. In all  simulations,
the results reported are for computations considering 100 gauge configurations rotated to the Landau gauge. 
Moreover, to increase the signal to noise ratio for all temperatures, with the exception of the highest,
the quark propagator was computed for two point sources located at $(0, 0, 0, 0)$ and $(L_t/2, 0, 0, 0)$ and the
results averaged before performing any analysis. For the highest temperature the
quark propagator was computed for a single source located at the origin of the lattice.

For  pure gauge theory the deconfinement temperature happens at $T_c = 270$ MeV, see e.g.~\cite{Silva:2016onh}
and references therein.
Our quark propagator simulations consider two temperatures below $T_c$, a temperature just above the
critical temperature and three temperatures clearly above $T_c$. In all cases, with the exception of two
temperatures, the quark propagator is computed for two bare quark masses $m_{bare} \approx 10$ and $50$ MeV.
In this way, one expects to reveal the fundamental properties of quark properties at finite temperature close
to the chiral limit. 

\section{The Lattice Propagator}\label{Sec:Prop}

\begin{figure*}[h] 
   \centering
   \includegraphics[width=6in]{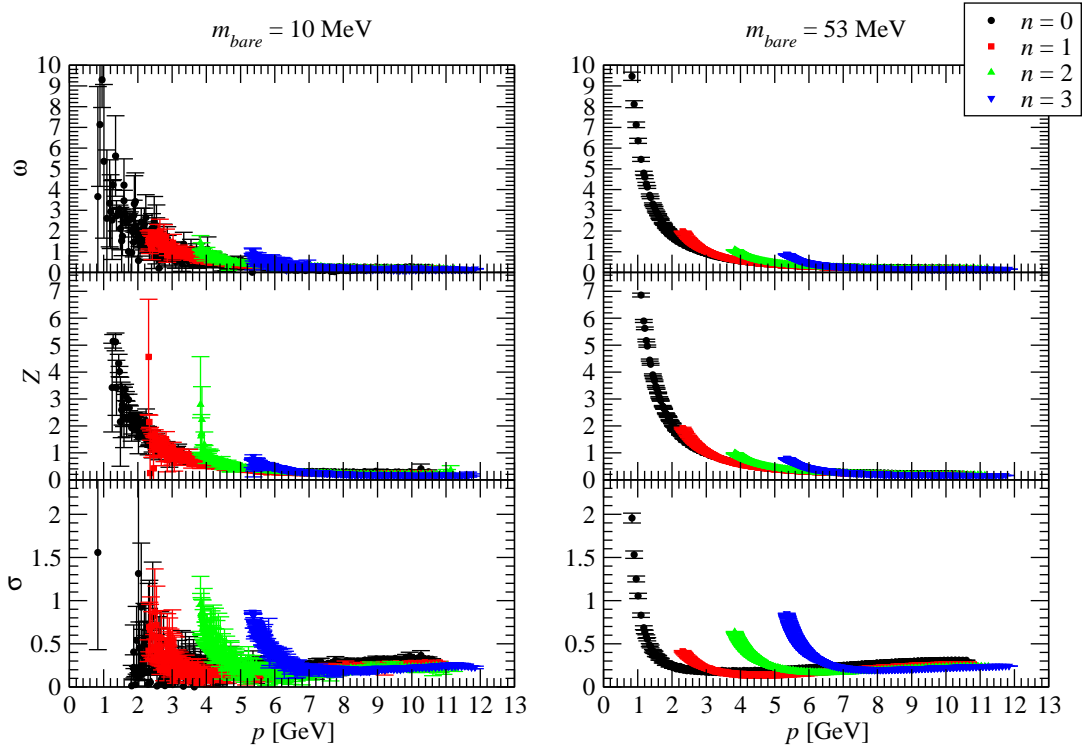} 
   \caption{The bare lattice form factors (\ref{EQ:LatOmega}), (\ref{EQ:LatZ}),  (\ref{EQ:LatM}), in lattice units,
                  as a function of  $p = \sqrt{ p^2_4 + \vec{p}^2}$ for the various Matsubara frequencies. 
                  The $m_{bare} = 11 MeV$ data for the smaller momenta has large fluctuations and some of the
                  data points fall outside the range of the axis values. In the legend on the $y$ axis, when reading $\omega$ it
                  refers to definition given in Eq. (\ref{EQ:LatOmega}), $Z$ to Eq. (\ref{EQ:LatZ}) and $\sigma$ to Eq. (\ref{EQ:LatM}).}
   \label{fig:fulllattff243}
\end{figure*}

\begin{figure*}[h] 
   \centering
   \includegraphics[width=6in]{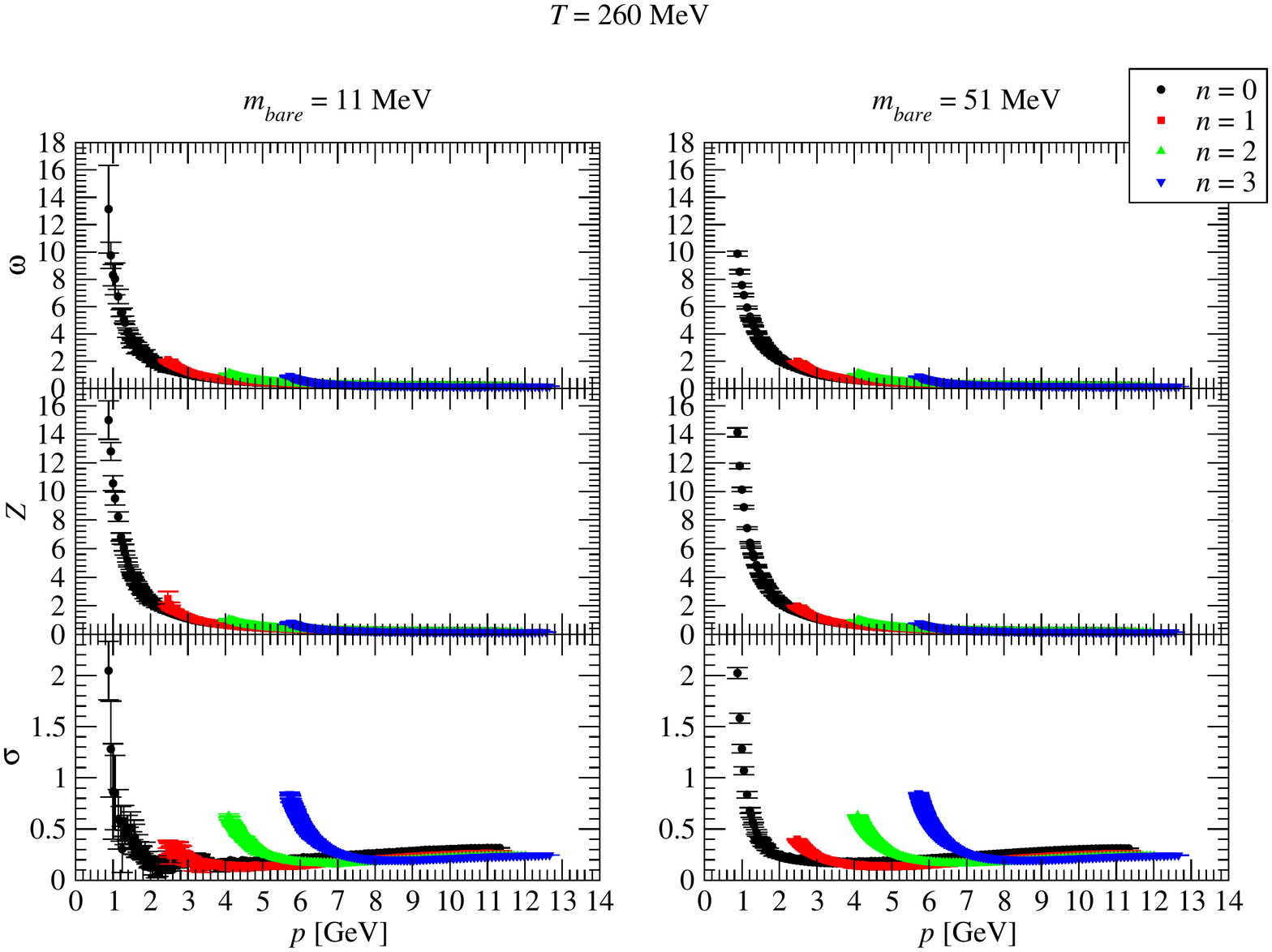} 
   \caption{The same as in Fig.~\ref{fig:fulllattff243} but for $T = 260$ MeV.}
   \label{fig:fulllattff260}
\end{figure*}

\begin{figure*}[h] 
   \centering
   \includegraphics[width=6in]{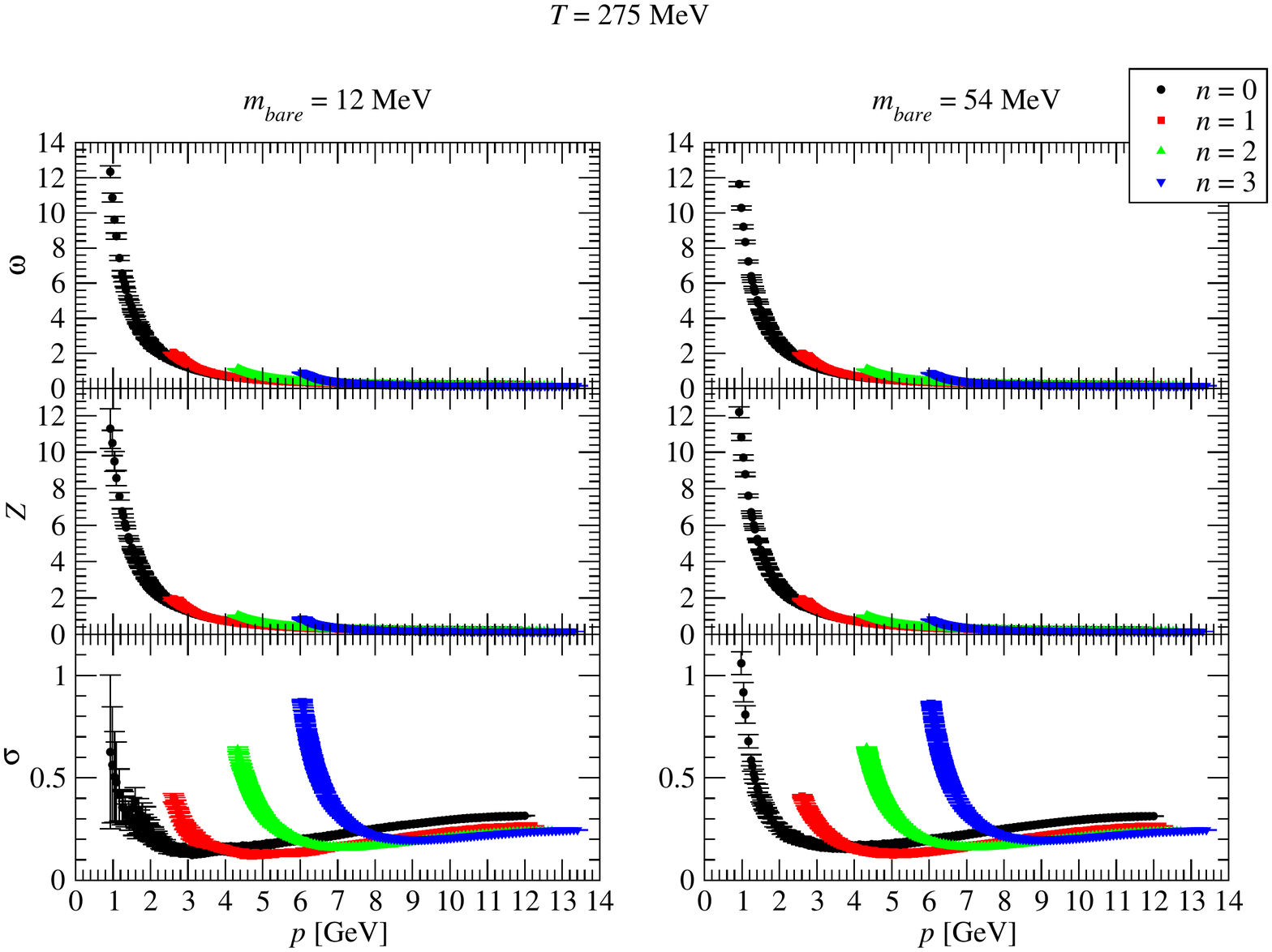} 
   \caption{The same as in Fig.~\ref{fig:fulllattff243} but for $T = 275$ MeV.}
   \label{fig:fulllattff275}
\end{figure*}

\begin{figure*}[h] 
   \centering
   \includegraphics[width=6in]{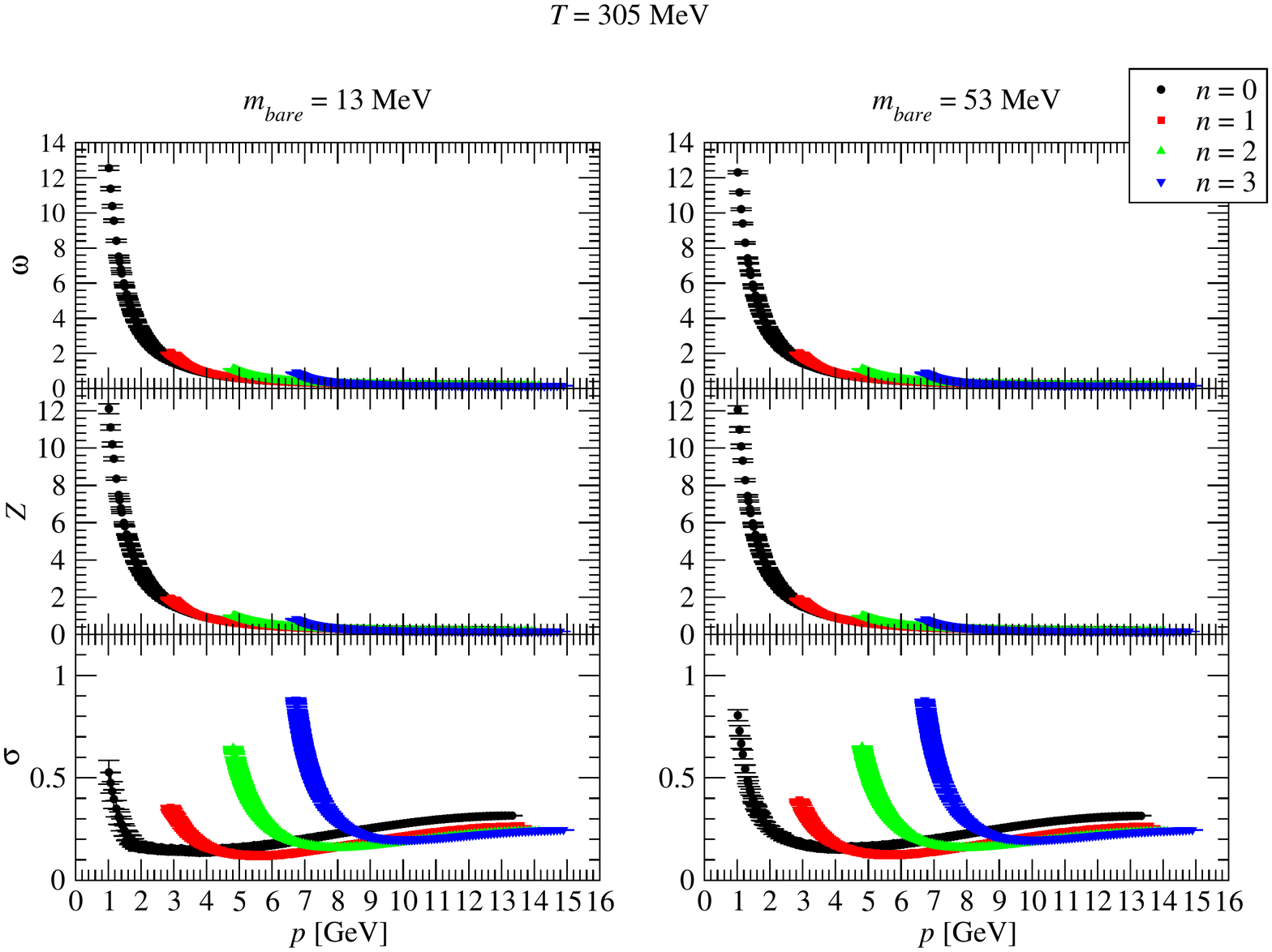} 
   \caption{The same as in Figs.~\ref{fig:fulllattff243}  but for $T = 305$ MeV.}
   \label{fig:fulllattff305}
\end{figure*}

The lattice form factors (\ref{EQ:LatOmega}), (\ref{EQ:LatZ}) and (\ref{EQ:LatM}) are measured by taking traces of 
the lattice propagator times gamma matrices. Their behaviour as a function of $p = \sqrt{p^2_4 + \vec{p}^2}$
is illustrated in Figs.~\ref{fig:fulllattff243}, ~\ref{fig:fulllattff260}, ~\ref{fig:fulllattff275} and ~\ref{fig:fulllattff305}  for
temperatures below and above the critical temperature $T_c = 270$ MeV and for all the Matsubara frequencies.

For the smallest bare quark mass and for the smallest temperature, the data shows large fluctuations but 
seems to follow the same pattern as the data for the heaviest bare quark mass.
For the smallest $m_{bare}$ and $T$, due to the large fluctuations observed in particular at the lower momenta,
some of data does not appear in the Fig. due our choice for the $y$ scale. 
Around and above $T_c$ the behaviour of the form factors for the two $m_{bare}$ is quite similar. 
The Figs. also show a clear violation of the rotational symmetry that occurs at all temperatures and
that is particularly dramatic for (\ref{EQ:LatM}).

\begin{figure*}[t] 
   \centering
   \includegraphics[width=6in]{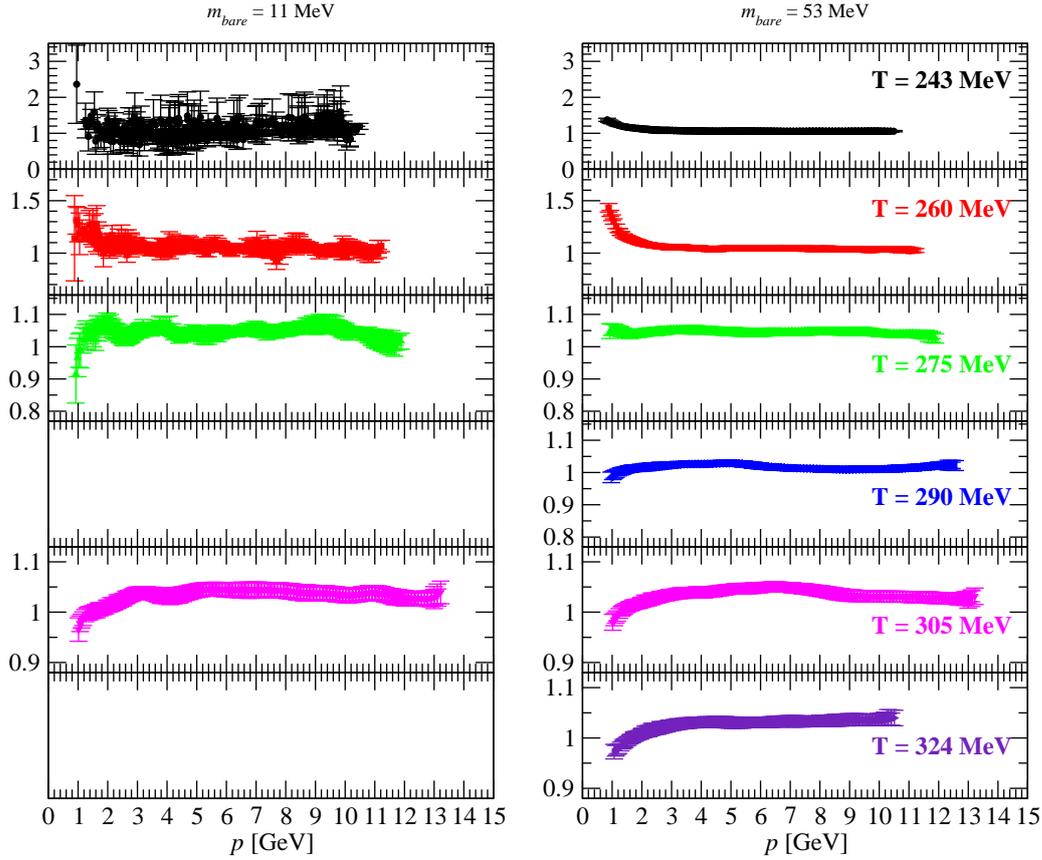}   
   \caption{Bare lattice quark wave function $Z_c(p_4, \vec{p})$ for the first Matsubara frequency as a function of $p = \sqrt{p^2_4 + \vec{p}^2}$.}
   \label{fig:Zc_Mat1}
\end{figure*}

\begin{figure*}[t] 
   \centering
   \includegraphics[width=6in]{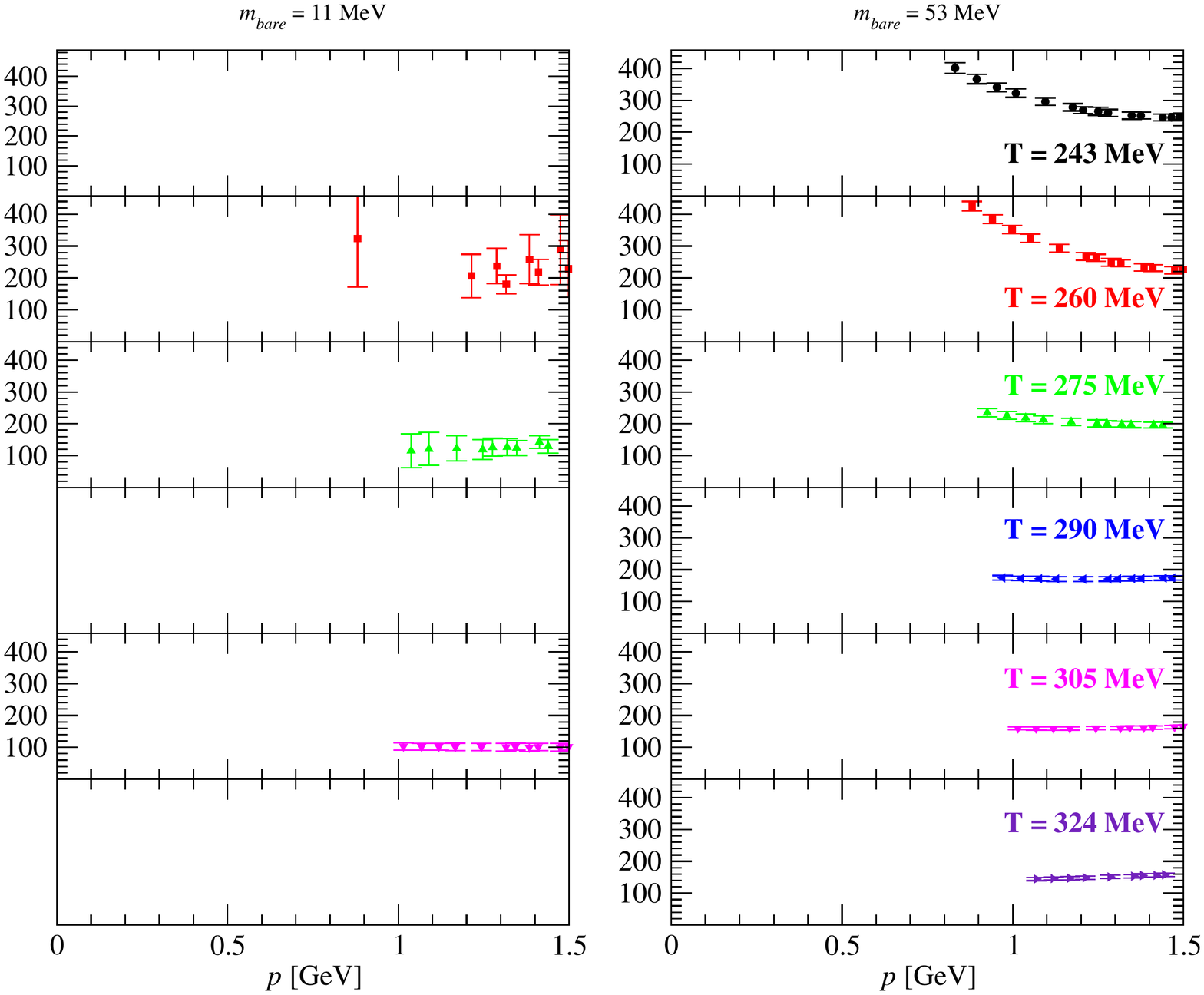}   
   \caption{Running quark mass, in GeV, for the first Matsubara frequency as a function of $p = \sqrt{p^2_4 + \vec{p}^2}$. Recall
   that for $Z_c$ and $M$, all data points whose relative error is greater than 50\% are ignored. This explains why some
   of the running mass plots are empty. It seems that $M$ has larger fluctuations when compared to the quark wave function.}
   \label{fig:M_Mat1}
\end{figure*}

The bare lattice quark wave function $Z_c (p_4, |\vec{p}|)$ for the full range of momenta, and the running quark mass $M (p_4, |\vec{p}|)$
for momenta up to 1.5 GeV can be seen on Figs.~\ref{fig:Zc_Mat1} and~\ref{fig:M_Mat1} for the first Matsubara frequency. 
Note that the data in the figures is not corrected for lattice spacing artefacts and that is the reason why for $M (p_4, |\vec{p}|)$ we only
report data for small $p = \sqrt{ p^2_4 + \vec{p}^2}$.
Indeed, the bare running quark mass at finite temperature looks like that observed at zero temperature
as seen, for example, in Fig. 10 in~\cite{Oliveira:2018lln}. This same Fig. also shows that up to momenta $a p \sim 0.5$
the lattice artefacts corrections are negligible or quite small. For the simulations reported here 
$a p \sim 0.5$ corresponds to a $p \sim 1$ GeV, see Tab.~\ref{Tab:LatSet}, and that is the reason why 
we plotted the running mass only up to $p = 1.5$ GeV. In this section, our analysis of the running quark mass
refers only to its low momentum behaviour. 
The function $M (p_4, |\vec{p}|)$ for the full range of momenta is discussed in Sec.~\ref{Sec:LattArt}.

As Figs.~\ref{fig:Zc_Mat1} and~\ref{fig:M_Mat1} show,
the quark propagator looks rather different below and above $T_c$.
For high momenta, the quark wave function approaches a constant from above for $T < T_c$ and from below at $T > T_c$.
Around the critical temperature the results favour a constant $Z_c$ for the full range of momenta accessed in our simulation. 
Fig~\ref{fig:Zc_Mat1} also shows that, for sufficiently high momenta, 
$\omega (p_4, \vec{p} ) \cong Z( p_4, \vec{p} )$, up to a constant factor close to unity.
 At low momenta $\omega (p_4, \vec{p} )$ exceeds $Z( p_4, \vec{p} )$
above the critical temperature, while $Z( p_4, \vec{p} )$ exceeds $\omega (p_4, \vec{p} )$ for temperatures
below $T_c$.
Note also that for temperatures above $T_c$  the quark wave function approaches a constant value from below,
reproducing the same type of behaviour observed for zero temperature and in agreement with the predictions of perturbation theory.

\begin{figure}[t] 
   \centering
   \includegraphics[width=3.5in]{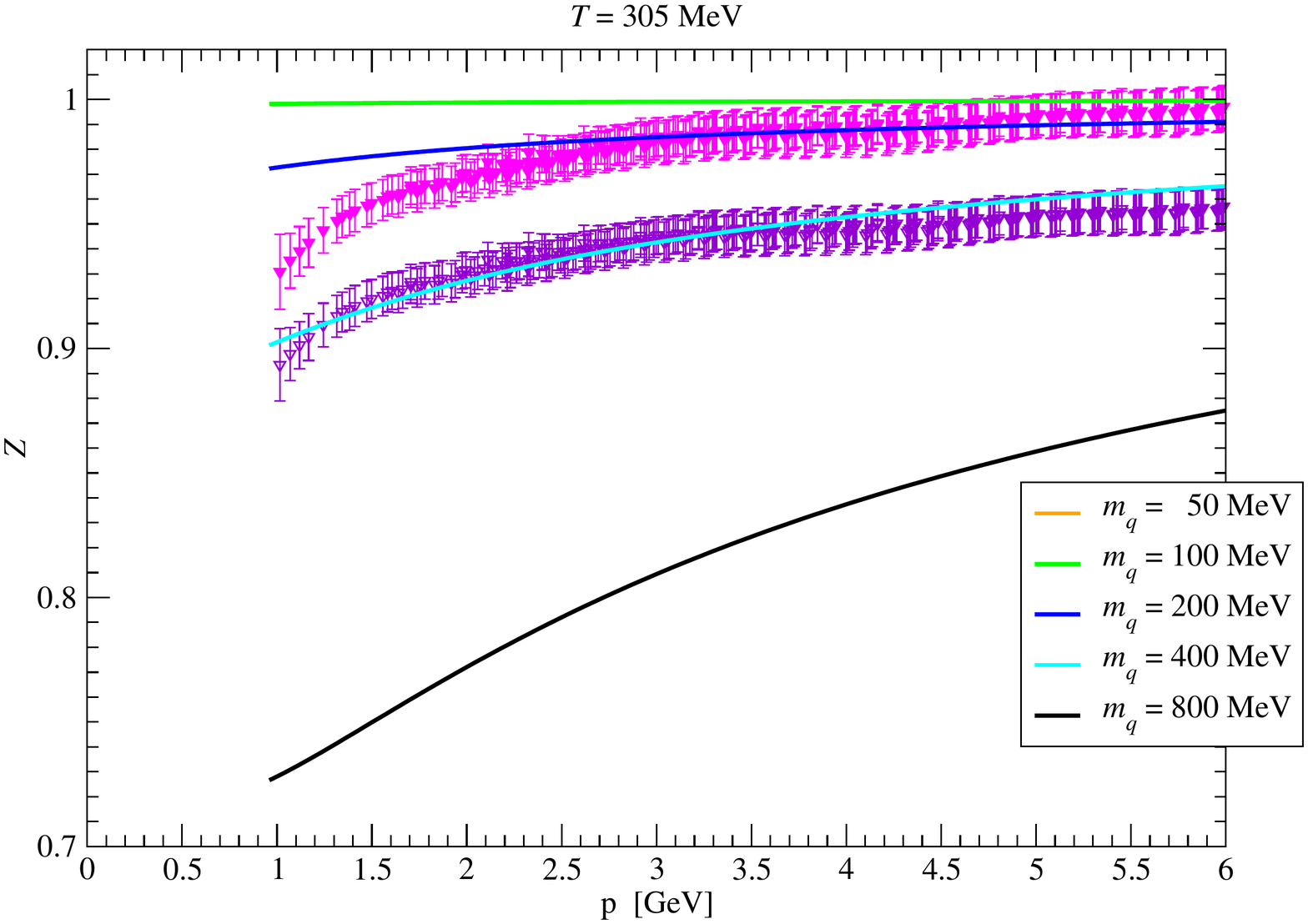} 
   \caption{The lattice and HTL quark wave function at $T = 305$ MeV. 
                  The lattice data was rescaled to reproduce the HTL wave function at 3 GeV for $m_q = 200$ MeV (full symbols) and
                  for $m_q = 400$ MeV (open symbols). Note that the data for $m_q = 50$ MeV and 100 MeV are indistinguishable in the plot.}
   \label{fig:Z_HTL_T305}
\end{figure}

\begin{figure}[t] 
   \centering
   \includegraphics[width=3.5in]{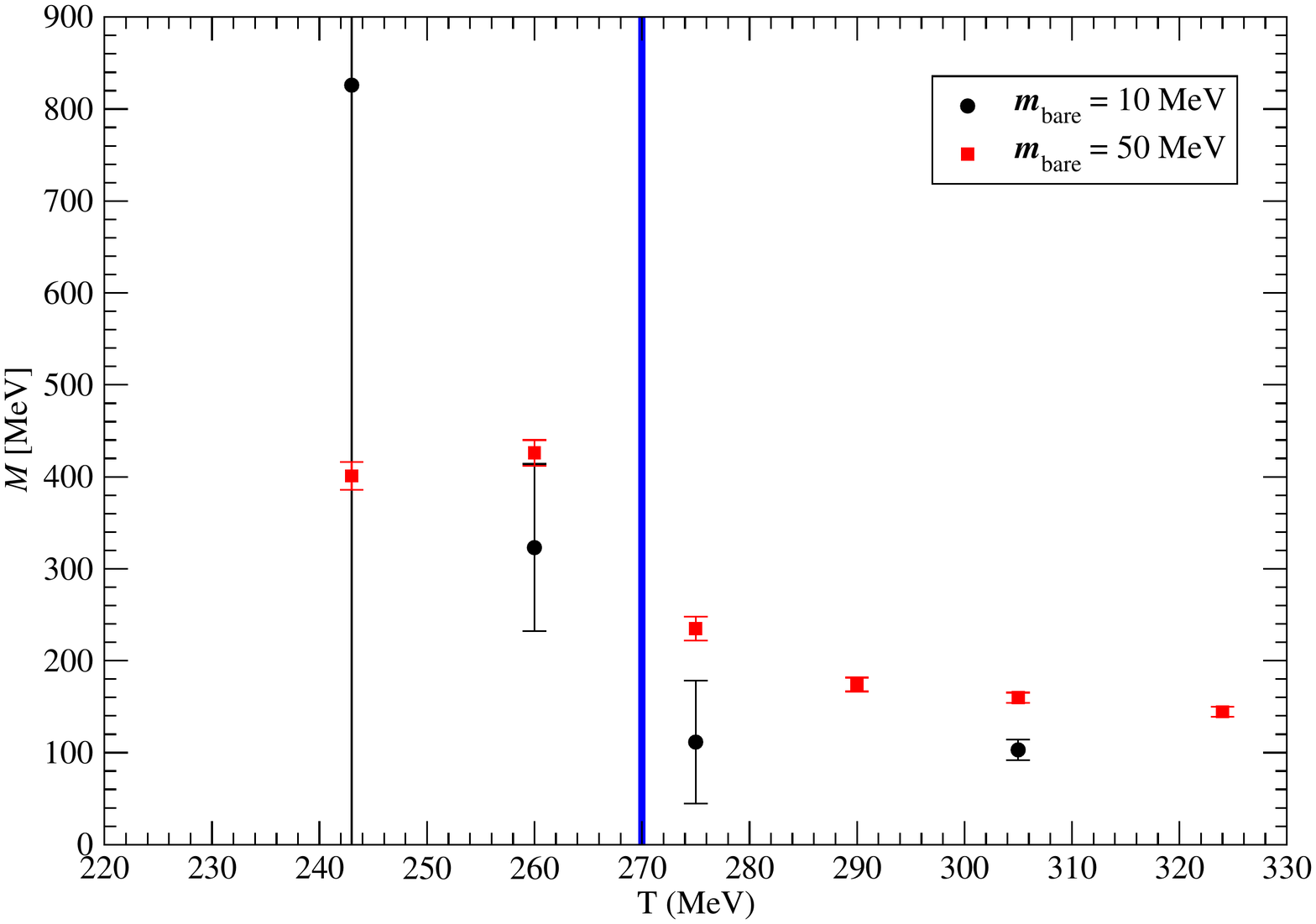} 
   \caption{Running quark mass at smaller momenta for the various $T$ considered in the current work.
                 the vertical blue line identifies the deconfinement transition.}
   \label{fig:M_T_0}
\end{figure}

In order to compare our results with the predictions of the hard thermal loop (HTL) for the quark propagator,
in Fig.~\ref{fig:Z_HTL_T305} we show the lattice data together with the one loop HTL prediction for $Z_c (p_4, \vec{p} )$ and for
$T = 305$ MeV, i.e. $T/T_c = 1.13$,  for the first Matsubara frequency. 
We take the  one loop HTL quark propagator from~\cite{Andersen:1999va}, and convert it to Euclidean space -- see their
Eqs. (5), (6) and (7).
The exact definition of the quark mass that should appear in the HTL expressions is difficult to 
read directly from the lattice data, see Figs.~\ref{fig:M_Mat1} and~\ref{fig:M_T_0}, and we have included in the plot for the
HTL quark wave function the corresponding numbers for a number of quark masses. 
Furthermore,  we also show twice the same lattice data rescaled to reproduce the HTL function at $p = 3$ GeV 
for $m_q = 200$ MeV and for $m_q = 400$ MeV. 
As can be observed, at higher momenta, i.e. for $p = ( p^2_4 + \vec{p}^2)^{1/2} \gtrsim 2$ GeV,
the lattice data is in good agreement with HTL.  
In particular for $m_q = 400$ MeV the lattice quark wave function and the HTL function are in good agreement for the full
range of momenta. 
Our conclusion being that for  $T / T_c = 1.13$, the lattice data is in good qualitative agreement with HTL for 
the quark wave function.

Similarly as for the quark wave function, the running quark mass shows different behaviours above and below the
deconfinement temperature (see Fig.~\ref{fig:M_Mat1}).
For temperatures below the critical temperature $M(p_4, \vec{p})$ is a decreasing function of
$p = \sqrt{p^2_4 + \vec{p}^2}$, while above the deconfinement phase transition it becomes essentially constant.
The observed behaviour of $M$ for $T > T_c$ supports the interpretation of quarks as quasiparticles with a constant mass.
Note, however, that for the largest temperature $M(p_4, \vec{p})$ seems to increase with $p$; see also the results reported in Sec.~\ref{Sec:LattArt}.
For $T > T_c$ the constant quark mass is a function of the temperature and, for the range of $T$ considered here, $M$ decreases
when $T$ increases.
This can be better viewed in Fig.~\ref{fig:M_T_0}, where $M(p_4, \vec{p} = 0)$ for the first Matsubara frequency
is shown for the full range of temperatures considered in the current simulations. 
The values of $M(p_4, \vec{p} = 0)$ reported in Fig.~\ref{fig:M_T_0} were computed using the data from the lowest momentum
that fulfill the cuts mentioned before and taking the data of the lattice form factors (\ref{EQ:LatOmega}), (\ref{EQ:LatZ}), (\ref{EQ:LatM}) 
with the errors computed assuming Gaussian error propagation. 
Note that the values reported in Fig.~\ref{fig:M_T_0} do not refer exactly to the same $p_4$, that ranges from 
just above $\sim 0.83$ GeV to just below $\sim 1.1$ GeV. 
It follows that for $T > T_c$ typical values for the mass of the quasiparticle are about $\sim 100$ MeV.

Although, our simulations are not at the chiral limit, the data for the running mass clearly shows a strong mass suppression 
as the temperature crosses $T_c$, with the values of $M$ given in Fig.~\ref{fig:M_T_0} above $T_c$ being about half of
the values reported when the quarks are in the confined phase.

\section{On the Correction of the Lattice Artefacts \label{Sec:LattArt}}

As discussed in~\cite{Skullerud:2000un,Skullerud:2001aw} for the 
zero temperature case, the lattice form factors computed directly from the inversion of the fermion matrix are 
contaminated by lattice artefacts. The solutions suggested to remove the lattice artefacts rely on the tree level lattice
quark propagator~\cite{Skullerud:2000un,Skullerud:2001aw}, on an expansion on the invariants of the H4 group~\cite{Becirevic:1999uc,deSoto:2007ht}
or on a combination of both methods~\cite{Oliveira:2018lln}. 

In what concerns the quark wave function at finite temperature, 
i.e. the results reported in Sec.~\ref{Sec:Prop} and summarized in Fig.~\ref{fig:Zc_Mat1},
the function $Z_c(p_4,\vec{p})$ looks rather flat for momenta above $\sim 2$ GeV for the temperatures investigated. 
This is precisely the type of functional behaviour predicted by perturbation theory. However,
these findings for $Z_c(p_4,\vec{p})$ contrast with the results of previous simulations at zero temperature, where for sufficiently high momenta
the lattice quark wave function is a decreasing function of $p$, even after the partial removing of the lattice artefacts 
based on the rotated tree level quark propagator~\cite{Skullerud:2000un,Skullerud:2001aw,Oliveira:2018lln}.
These results follow because $Z_c(p_4,\vec{p})$ is measured as a ratio of functions that, according to the
procedure mentioned above, have exactly the same type of lattice artefacts corrections and, therefore, by taking
ratios of these type of functions the lattice artefacts cancel exactly or, at least, are strongly suppressed. 
Given the good agreement between the lattice $Z_c(p_4,\vec{p})$ reported in Fig.~\ref{fig:Zc_Mat1} and the results of perturbation
theory for $p \gtrsim 2$ GeV, we assume that the computed $Z_c(p_4,\vec{p})$ are essentially free of lattice artefacts.

On the other hand, the running quark mass at finite temperature is computed as in the zero temperature case. 
Not surprisingly, the lattice data for $M(p_4, \vec{p})$ shows a similar pattern as observed at zero temperature and 
this function increases at higher momenta as observed in Figs. 1 and 10 of~\cite{Oliveira:2018lln} for the uncorrected lattice data.
It follows that the computation of the running quark mass requires an estimation and subtraction of the lattice artefacts.

\begin{figure}[t] 
   \centering
   \includegraphics[width=3.5in]{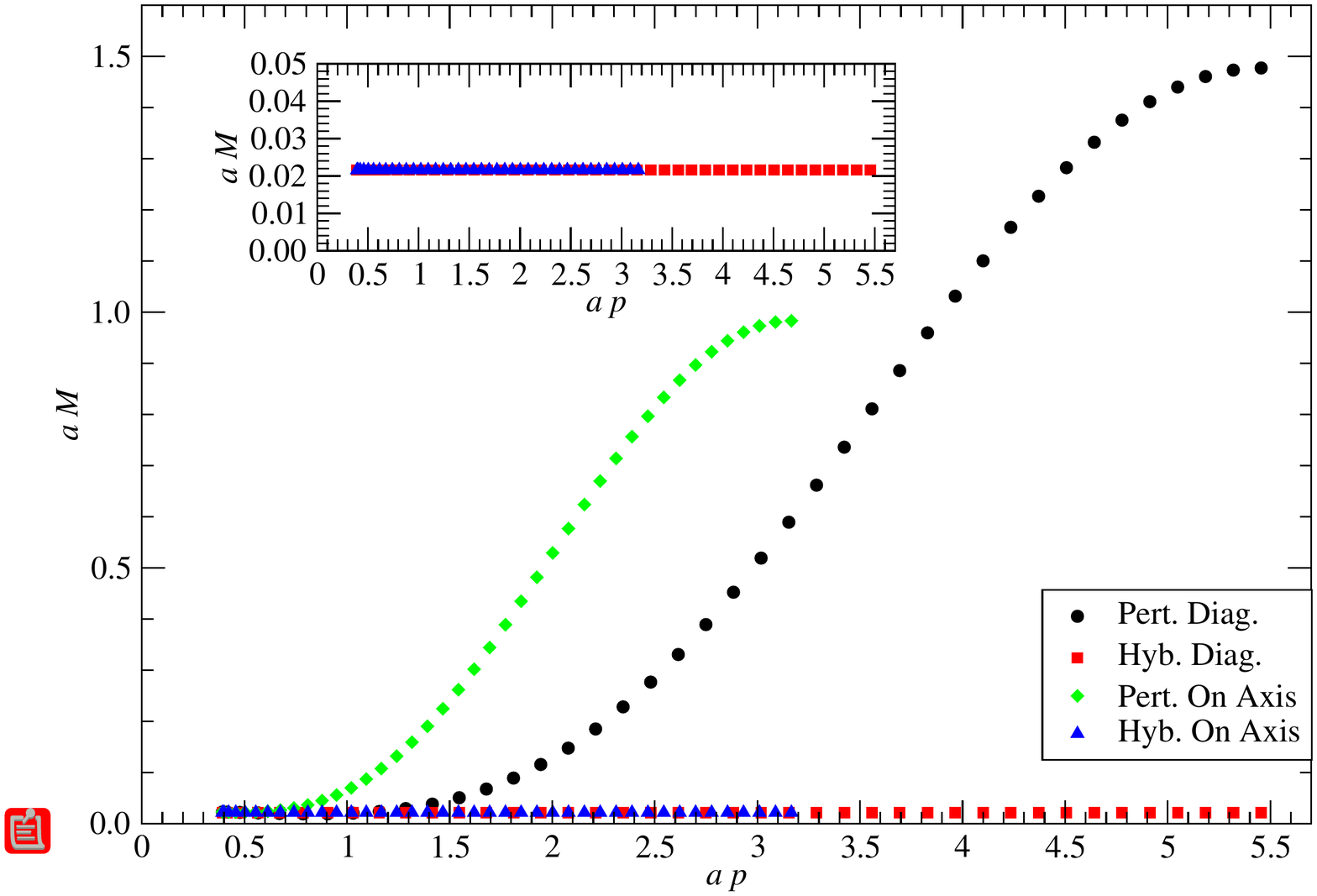}
   \caption{Tree level propagator mass for three momenta $\vec{p}$ being along the diagonal of the axis of the lattice.
                 The data reported was computed for $T = 305$ MeV on a $80^3 \times 8$ lattice, $\kappa = 0.1348$ that corresponds to $m_{bare} =
                 53$ MeV, and for the first Matsubara frequency.
                 The data referred as ``Pert.'' is the na\"{\i}ve tree level pole mass appearing on the rotated quark propagator.
                 The data labelled as "Hyb.'' is the tree level corrected result for the so-called hybrid definition of the mass.
                 See~\cite{Skullerud:2000un,Skullerud:2001aw,Oliveira:2018lln} for details.}
   \label{fig:MPertCorr}
\end{figure}

For the estimation of the lattice artefacts for the running mass we use the procedure outlined 
in~\cite{Skullerud:2000un,Skullerud:2001aw}.
The space-spin components of the rotated tree-level quark propagator is given by
\begin{equation}
S^{(0)}(p) = 
\frac{ -i \, a \slashed{\overline p} ~ A(m, p) + B (m, p)}{a^2 \overline p^2 A^2(m, p) + B^2(m, p)} \ .
\label{Eq:treelevelS}
\end{equation}
Expressions for $A(m, p)$ and $B (m, p)$ can be found in~\cite{Skullerud:2000un,Skullerud:2001aw}.
There is no unambiguous way to subtract the lattice artefacts. Typically, two main definitions of the ``continuum'' running
mass are considered where one na\"{\i}vely subtracts  part of the lattice artefacts, defining $m_s ( p_4, \vec{p})$,
and the so-called hybrid scheme that defines $m_h ( p_4, \vec{p})$,
where in the subtraction of the lattice artefacts it is taking into consideration if the corrections give either positive or negative 
contributions to the pole mass. 
In the definition of $m_h ( p_4, \vec{p})$ the negative contributions are subtracted before rescaling
the result to take into account the remaining corrections. 
Details of the procedure and definitions can be found in the above cited works.
To illustrate the running quark mass definition, on Fig.~\ref{fig:MPertCorr} we report results obtained
with the rotated tree level quark propagator, where the na\"{\i}ve pole mass grows with momenta and $m_h ( p_4, \vec{p})$
is constant for all $a p$.

\begin{figure}[t] 
   \centering
   \includegraphics[width=3.5in]{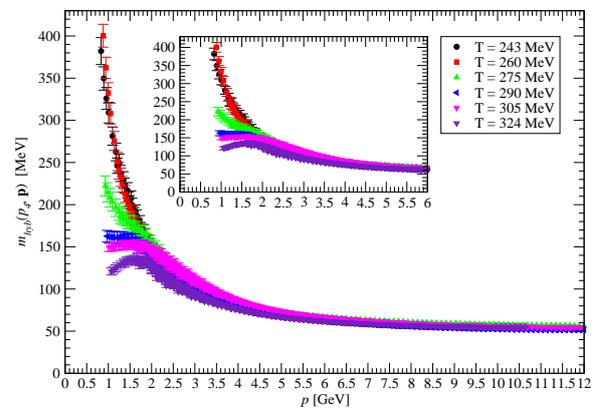}
   \caption{The running quark mass $m_h ( p_4, \vec{p})$ for all the $T$ and for the simulations using $m_{bare} \sim 50$ MeV.
                 Only the data corresponding to the first Matsubara is shown.}
   \label{fig:Mhyb50}
\end{figure}

\begin{figure}[t] 
   \centering
   \includegraphics[width=3.5in]{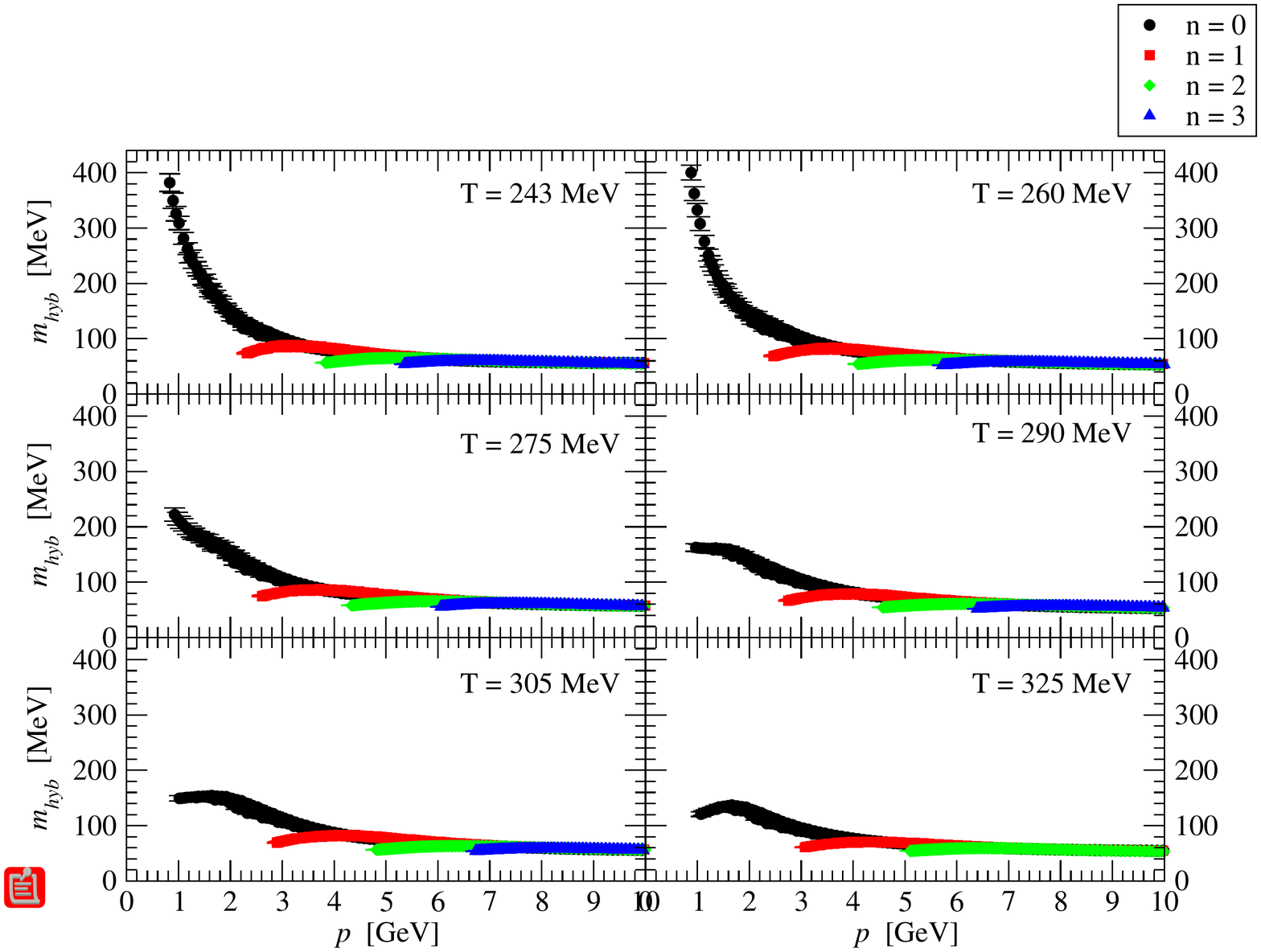}
   \caption{The running quark mass $m_h ( p_4, \vec{p})$ for all the $T$ and for all Matsubara frequencies computed for the simulations 
                   using $m_{bare} \sim 50$ MeV.}
   \label{fig:Mhyb50All}
\end{figure}

The hybrid running quark mass for the simulations with a $m_{bare} \approx 50$ MeV and all temperatures are given in
Fig.~\ref{fig:Mhyb50} for the first Matsubara frequency. The functional form at small momenta for $m_h( p_4, \vec{p})$ 
changes dramatically when the deconfinement transition is approached. At the  infrared scales $m_h$ is a decreasing function
of $p$ for $T \lesssim T_c$, is essentially constant for $T$ above $T_c$ and in the range 290 - 305 MeV and for the highest  temperature
considered here $m_h$ is an increasing function of the momenta. We call the reader's attention that in the simulation for $T = 324$ MeV the
number of lattice points in the time direction, used to define the temperature, is shorter $L_t = 6$ compared with all other cases that use
$L_t = 8$. It is not clear if such a short time direction is responsible for the different observed behaviour on the hybrid running quark mass.
At sufficiently high $p$, the data shows that the running masses seem to collapse into a single curve. 
The data reported on Fig.~\ref{fig:Mhyb50} suggests that at high $p$ and within the range of momenta accessed in our simulation
$m_h(p_4, \vec{p})$ is a decreasing function of $p$. We have tried to fit a constant value to the highest momenta and have verified that
for $p \gtrsim 10$ GeV the lattice data for $m_h(p_4, \vec{p})$ is compatible with a constant just above 50 MeV; for example,
for $T = 290$ MeV, we get  $m_h = 51.984 \pm 0.030$ MeV with a $\chi^2/d.o.f = 1.1$ for $p > 9.8$ GeV. This fit value for the mass 
is within the same range of values for $m_{bare}$ considered here -- see Tab.~\ref{Tab:LatSet}. 

For completeness, in Fig.~\ref{fig:Mhyb50All} we show $m_h(p_4, \vec{p})$ for all temperatures and all the Matsubara frequencies.
The data shows, once more, a clear violation of  rotational symmetry.

The computed quark wave function $Z_c(p_4, \vec{p})$ and running quark mass $m_h(p_4, \vec{p})$ imply changes on the 
functional behaviour of the quark spectral functions below and above the critical temperature. However, given the small number of
Matsubara frequencies accessed in our study, we do not attempt to compute any of the components of the quark spectral function.

\section{Summary and Conclusions}

In the current work we report on the computation of the finite temperature Landau gauge lattice QCD quark propagator in the
quenched approximation at small quark masses $\sim 10$ MeV and $\sim 50$ MeV for temperatures below and above
the deconfinement phase transition.
The various quark form factors are investigated as a function of the temperature for the various Matsubara frequencies. 

Our results show that both the quark wave function and the running quark mass have different functional forms for 
$T < T_c$ and for $T > T_c$. Above $T_c$ the quark wave function seems to reproduce the predictions of HTL expansion 
for $T / T_c  \gtrsim 1.1$ being constant for $p \gtrsim 2$ GeV and slightly suppressed at smaller momenta. The suppression
of the quark wave function at infrared scales is also observed for the zero temperature case; see~\cite{Oliveira:2018lln} and references 
therein.
On the other hand,  at low momenta and for $T < T_c$ the quark wave function is enhanced. These results
show that $Z_c$ has a non-trivial dependence on the temperature and its suppression or enhancement at low momenta is not
enough to identify if the quark is on the confined or deconfined phase.

The running quark mass also shows a non-trivial dependence with the temperature of the heat bath. Indeed, the
running quark mass is highly suppressed when $T$ crosses $T_c$ from below,
with typical values just above $T_c$ being about half of the corresponding values just below $T_c$.
At low momenta $p \lesssim 1.5$ GeV for the temperatures considered here,
the running quark mass is nearly constant above $T_c$.
This can be viewed as favouring the description of quarks as quasi free particles with a constant mass, as is used e.g. by the HTL approach.
Our simulations also show that at high momenta the running quark mass reproduces the values of the bare quark mass 
reported in Tab.~\ref{Tab:LatSet}. 
For temperatures below $T_c$ the running quark mass increase when one approaches the zero momentum limit, reproducing
the observed behaviour for the running quark mass at zero temperature.

Our simulations are based on pure gauge configurations, 
where the pure gauge sector is solved exactly but without taking into account the dynamics of quarks.
The computed quark propagators refer to the propagation of the quark fields on a background of gluons that is solved exactly for
the bosonic sector of the theory. The  observed significant suppression of the running quark mass for temperatures above $T_c$
is an indication of a major contribution of the gluons for the mechanism of chiral symmetry breaking. Furthermore, our results
suggest an important link between (gluon) confinement and the restoration of chiral symmetry that needs to be further investigated.

\begin{acknowledgements}
The authors acknowledge M. Strickland and P. Costa for helpful discussions.
The authors acknowledge financial support from FCT under contract with reference UID/FIS/04564/2016. 
P.J.S. also acknowledges partial support by Funda\c{c}\~ao para a Ci\^encia e a Tecnologia (FCT) under contracts 
SFRH/BPD/40998/2007 and SFRH/BPD/109971/2015. The authors also acknowledge the Laboratory for Advanced 
Computing at University of Coimbra (\url{http://www.uc.pt/lca}) for providing access to the HPC resource Navigator. 
The SU(3) lattice simulations were done using Chroma \cite{Edwards2005} and PFFT \cite{Pippig2013} libraries.

\end{acknowledgements}


\begin{thebibliography}{99}


\bibitem{LeBellac:1991cq}
  M.~Le Bellac,
  ``Thermal Field Fheory,''
  Cambridge, UK: CUP (2000)
  
\bibitem{Kapusta:2006pm}
  J.~I.~Kapusta and C.~Gale,
  ``Finite-temperature field theory: Principles and applications,'

\bibitem{Su:2015esa}
  N.~Su,
  Int.\ J.\ Mod.\ Phys.\ A {\bf 30} (2015) 1530025
  doi:10.1142/S0217751X15300252
  [arXiv:1502.04589 [hep-ph]].
  
  
  
  \bibitem{McLerran:1980pk}
  L.~D.~McLerran and B.~Svetitsky,
  Phys.\ Lett.\  {\bf 98B} (1981) 195.
  doi:10.1016/0370-2693(81)90986-2
  
\bibitem{McLerran:1981pb}
  L.~D.~McLerran and B.~Svetitsky,
  Phys.\ Rev.\ D {\bf 24} (1981) 450.
  doi:10.1103/PhysRevD.24.450
  

\bibitem{Silva:2016onh}
  P.~J.~Silva and O.~Oliveira,
  Phys.\ Rev.\ D {\bf 93} (2016) no.11,  114509
  doi:10.1103/PhysRevD.93.114509
  [arXiv:1601.01594 [hep-lat]].  

    


\bibitem{Aouane:2011fv}
  R.~Aouane, V.~G.~Bornyakov, E.~M.~Ilgenfritz, V.~K.~Mitrjushkin, M.~Muller-Preussker and A.~Sternbeck,
  Phys.\ Rev.\ D {\bf 85} (2012) 034501
  doi:10.1103/PhysRevD.85.034501
  [arXiv:1108.1735 [hep-lat]].



\bibitem{Silva:2013maa}
  P.~J.~Silva, O.~Oliveira, P.~Bicudo and N.~Cardoso,
  Phys.\ Rev.\ D {\bf 89} (2014) no.7,  074503
  doi:10.1103/PhysRevD.89.074503
  [arXiv:1310.5629 [hep-lat]].
  
\bibitem{Huber:2018ned}
For a recent review at zero temperature see e.g.
  M.~Q.~Huber,
  arXiv:1808.05227 [hep-ph].  

\bibitem{Aouane:2012bk}
  R.~Aouane, F.~Burger, E.-M.~Ilgenfritz, M.~Müller-Preussker and A.~Sternbeck,
  Phys.\ Rev.\ D {\bf 87} (2013) no.11,  114502
  doi:10.1103/PhysRevD.87.114502
  [arXiv:1212.1102 [hep-lat]].


\bibitem{Silva:2016msq}
  P.~J.~Silva, O.~Oliveira, D.~Dudal, P.~Bicudo and N.~Cardoso,
  Few Body Syst.\  {\bf 58} (2017) no.3,  127
  doi:10.1007/s00601-017-1281-7
  [arXiv:1611.04966 [hep-lat]].
  
  
\bibitem{Maas:2011se}
  A.~Maas,
  Phys.\ Rept.\  {\bf 524} (2013) 203
  doi:10.1016/j.physrep.2012.11.002
  [arXiv:1106.3942 [hep-ph]].  
  
  
      
    
    

\bibitem{Ikeda:2001vc}
  T.~Ikeda,
  Prog.\ Theor.\ Phys.\  {\bf 107} (2002) 403
  doi:10.1143/PTP.107.403
  [hep-ph/0107105].
  
\bibitem{Mueller:2010ah}
  J.~A.~Mueller, C.~S.~Fischer and D.~Nickel,
  Eur.\ Phys.\ J.\ C {\bf 70} (2010) 1037
  doi:10.1140/epjc/s10052-010-1499-8
  [arXiv:1009.3762 [hep-ph]].
  
\bibitem{Fischer:2010fx}
  C.~S.~Fischer, A.~Maas and J.~A.~Muller,
  Eur.\ Phys.\ J.\ C {\bf 68} (2010) 165
  doi:10.1140/epjc/s10052-010-1343-1
  [arXiv:1003.1960 [hep-ph]].
  
\bibitem{Contant:2017gtz}
  R.~Contant and M.~Q.~Huber,
  Phys.\ Rev.\ D {\bf 96} (2017) no.7,  074002
  doi:10.1103/PhysRevD.96.074002
  [arXiv:1706.00943 [hep-ph]].  
  
\bibitem{Fischer:2009wc}
  C.~S.~Fischer,
  Phys.\ Rev.\ Lett.\  {\bf 103} (2009) 052003
  doi:10.1103/PhysRevLett.103.052003
  [arXiv:0904.2700 [hep-ph]].  
  
\bibitem{Qin:2010pc}
  S.~x.~Qin, L.~Chang, Y.~x.~Liu and C.~D.~Roberts,
  Phys.\ Rev.\ D {\bf 84} (2011) 014017
  doi:10.1103/PhysRevD.84.014017
  [arXiv:1010.4231 [nucl-th]].  

\bibitem{Weldon:1999th}
  H.~A.~Weldon,
  Phys.\ Rev.\ D {\bf 61} (2000) 036003
  doi:10.1103/PhysRevD.61.036003
  [hep-ph/9908204].


\bibitem{Nakkagawa:2011ci}
  H.~Nakkagawa, H.~Yokota and K.~Yoshida,
  Phys.\ Rev.\ D {\bf 85} (2012) 031902
  doi:10.1103/PhysRevD.85.031902
  [arXiv:1111.0117 [hep-ph]].
  
\bibitem{Nakkagawa:2012ip}
  H.~Nakkagawa, H.~Yokota and K.~Yoshida,
  Phys.\ Rev.\ D {\bf 86} (2012) 096007
  doi:10.1103/PhysRevD.86.096007
  [arXiv:1208.6386 [hep-ph]].  



\bibitem{Harada:2008vk}
  M.~Harada and Y.~Nemoto,
  Phys.\ Rev.\ D {\bf 78} (2008) 014004
  doi:10.1103/PhysRevD.78.014004
  [arXiv:0803.3257 [hep-ph]].



\bibitem{Satow:2010ia}
  D.~Satow, Y.~Hidaka and T.~Kunihiro,
  Phys.\ Rev.\ D {\bf 83} (2011) 045017
  doi:10.1103/PhysRevD.83.045017
  [arXiv:1011.6452 [hep-ph]].
  
  
\bibitem{Hidaka:2011rz}
  Y.~Hidaka, D.~Satow and T.~Kunihiro,
  Nucl.\ Phys.\ A {\bf 876} (2012) 93
  doi:10.1016/j.nuclphysa.2011.12.007
  [arXiv:1111.5015 [hep-ph]].  
  
\bibitem{Kitazawa:2005mp}
  M.~Kitazawa, T.~Kunihiro and Y.~Nemoto,
  Phys.\ Lett.\ B {\bf 633} (2006) 269
  doi:10.1016/j.physletb.2005.11.076
  [hep-ph/0510167].
  
 \bibitem{Kitazawa:2006zi}
  M.~Kitazawa, T.~Kunihiro and Y.~Nemoto,
  Prog.\ Theor.\ Phys.\  {\bf 117} (2007) 103
  doi:10.1143/PTP.117.103
  [hep-ph/0609164].
  
\bibitem{Kitazawa:2006vh}
  M.~Kitazawa, T.~Kunihiro and Y.~Nemoto,
  Nucl.\ Phys.\ A {\bf 785} (2007) 257
  doi:10.1016/j.nuclphysa.2006.11.077
  [hep-ph/0608185].
  
\bibitem{Kitazawa:2007ep}
  M.~Kitazawa, T.~Kunihiro, K.~Mitsutani and Y.~Nemoto,
  Phys.\ Rev.\ D {\bf 77} (2008) 045034
  doi:10.1103/PhysRevD.77.045034
  [arXiv:0710.5809 [hep-ph]].    


\bibitem{Wang:2018osm}
  Z.~Wang and L.~He,
  Phys.\ Rev.\ D {\bf 98} (2018) no.9,  094031
  doi:10.1103/PhysRevD.98.094031
  [arXiv:1808.08535 [hep-ph]].






\bibitem{Hamada:2006ra}
  M.~Hamada, H.~Kouno, A.~Nakamura, T.~Saito and M.~Yahiro,
  PoS LAT {\bf 2006} (2006) 136
  doi:10.22323/1.032.0136
  [hep-lat/0610010].
  
\bibitem{Karsch:2007wc}
  F.~Karsch and M.~Kitazawa,
  Phys.\ Lett.\ B {\bf 658} (2007) 45
  doi:10.1016/j.physletb.2007.10.034
  [arXiv:0708.0299 [hep-lat]].
  
\bibitem{Karsch:2007bf}
  F.~Karsch and M.~Kitazawa,
  PoS LATTICE {\bf 2007} (2007) 197
  doi:10.22323/1.042.0197
  [arXiv:0710.2948 [hep-lat]].    
  
\bibitem{Hamada:2008zz}
  M.~Hamada, M.~Yahiro, H.~Kouno, A.~Nakamura and T.~Saito,
  PoS LATTICE {\bf 2008} (2008) 210.
  doi:10.22323/1.066.0210

\bibitem{Karsch:2009tp}
  F.~Karsch and M.~Kitazawa,
  Phys.\ Rev.\ D {\bf 80} (2009) 056001
  doi:10.1103/PhysRevD.80.056001
  [arXiv:0906.3941 [hep-lat]].

\bibitem{Kitazawa:2009uw}
  M.~Kitazawa and F.~Karsch,
  Nucl.\ Phys.\ A {\bf 830} (2009) 223C
  doi:10.1016/j.nuclphysa.2009.09.024
  [arXiv:0908.3079 [hep-lat]].
  
\bibitem{Hamada:2010zz}
  M.~Hamada, H.~Kouno, A.~Nakamura, T.~Saito and M.~Yahiro,
  Phys.\ Rev.\ D {\bf 81} (2010) 094506.
  doi:10.1103/PhysRevD.81.094506

\bibitem{Kaczmarek:2012mb}
  O.~Kaczmarek, F.~Karsch, M.~Kitazawa and W.~Soldner,
  Phys.\ Rev.\ D {\bf 86} (2012) 036006
  doi:10.1103/PhysRevD.86.036006
  [arXiv:1206.1991 [hep-lat]].


\bibitem{Oliveira:2012eh}
  O.~Oliveira and P.~J.~Silva,
  Phys.\ Rev.\ D {\bf 86} (2012) 114513
  doi:10.1103/PhysRevD.86.114513
  [arXiv:1207.3029 [hep-lat]].
  
\bibitem{Duarte:2016iko}
  A.~G.~Duarte, O.~Oliveira and P.~J.~Silva,
  Phys.\ Rev.\ D {\bf 94} (2016) no.1,  014502
  doi:10.1103/PhysRevD.94.014502
  [arXiv:1605.00594 [hep-lat]].
  
\bibitem{Dudal:2018cli}
  D.~Dudal, O.~Oliveira and P.~J.~Silva,
  Annals Phys.\  {\bf 397} (2018) 351
  doi:10.1016/j.aop.2018.08.019
  [arXiv:1803.02281 [hep-lat]].
  
\bibitem{Sheikholeslami:1985ij}
  B.~Sheikholeslami and R.~Wohlert,
  Nucl.\ Phys.\ B {\bf 259} (1985) 572.
  doi:10.1016/0550-3213(85)90002-1  
  
\bibitem{Luscher:1996sc}
  M.~Luscher, S.~Sint, R.~Sommer and P.~Weisz,
  Nucl.\ Phys.\ B {\bf 478} (1996) 365
  doi:10.1016/0550-3213(96)00378-1
  [hep-lat/9605038].  
  
\bibitem{Heatlie:1990kg}
  G.~Heatlie, G.~Martinelli, C.~Pittori, G.~C.~Rossi and C.~T.~Sachrajda,
  Nucl.\ Phys.\ B {\bf 352} (1991) 266.
  doi:10.1016/0550-3213(91)90137-M  

\bibitem{Skullerud:2000un}
  J.~I.~Skullerud and A.~G.~Williams,
  Phys.\ Rev.\ D {\bf 63} (2001) 054508
  doi:10.1103/PhysRevD.63.054508
  [hep-lat/0007028].
  
\bibitem{Skullerud:2001aw}
  J.~Skullerud, D.~B.~Leinweber and A.~G.~Williams,
  Phys.\ Rev.\ D {\bf 64} (2001) 074508
  doi:10.1103/PhysRevD.64.074508
  [hep-lat/0102013].
  
\bibitem{Luscher:1996ug}
  M.~Luscher, S.~Sint, R.~Sommer, P.~Weisz and U.~Wolff,
  Nucl.\ Phys.\ B {\bf 491} (1997) 323
  doi:10.1016/S0550-3213(97)00080-1
  [hep-lat/9609035].  


  
\bibitem{Leinweber:1998uu}
  D.~B.~Leinweber {\it et al.} [UKQCD Collaboration],
  Phys.\ Rev.\ D {\bf 60} (1999) 094507
   Erratum: [Phys.\ Rev.\ D {\bf 61} (2000) 079901]
  doi:10.1103/PhysRevD.61.079901, 10.1103/PhysRevD.60.094507
  [hep-lat/9811027].
    
  
\bibitem{Oliveira:2018lln}
  O.~Oliveira, P.~J.~Silva, J.~I.~Skullerud and A.~Sternbeck,
  arXiv:1809.02541 [hep-lat].  
  
\bibitem{Andersen:1999va}
  J.~O.~Andersen, E.~Braaten and M.~Strickland,
  Phys.\ Rev.\ D {\bf 61} (2000) 074016
  doi:10.1103/PhysRevD.61.074016
  [hep-ph/9908323].  
  
\bibitem{Skullerud:2001aw}
  J.~Skullerud, D.~B.~Leinweber and A.~G.~Williams,
  Phys.\ Rev.\ D {\bf 64} (2001) 074508
  doi:10.1103/PhysRevD.64.074508
  [hep-lat/0102013].  
  
  
\bibitem{Boucaud:2017ksi}
  P.~Boucaud, F.~De Soto, J.~Rodr\'{\i}guez-Quintero and S.~Zafeiropoulos,
  Phys.\ Rev.\ D {\bf 96} (2017) no.9,  098501
  doi:10.1103/PhysRevD.96.098501
  [arXiv:1704.02053 [hep-lat]].
  
  \bibitem{Duarte:2017wte}
  A.~G.~Duarte, O.~Oliveira and P.~J.~Silva,
  Phys.\ Rev.\ D {\bf 96} (2017) no.9,  098502
  doi:10.1103/PhysRevD.96.098502
  [arXiv:1704.02864 [hep-lat]].  
  
  

\bibitem{Becirevic:1999uc}
  D.~Becirevic, P.~Boucaud, J.~P.~Leroy, J.~Micheli, O.~Pene, J.~Rodriguez-Quintero and C.~Roiesnel,
  Phys.\ Rev.\ D {\bf 60} (1999) 094509
  doi:10.1103/PhysRevD.60.094509
  [hep-ph/9903364].
  
\bibitem{deSoto:2007ht}
  F.~de Soto and C.~Roiesnel,
  JHEP {\bf 0709} (2007) 007
  doi:10.1088/1126-6708/2007/09/007
  [arXiv:0705.3523 [hep-lat]].
  
\bibitem{Edwards2005}
R. G. Edwards, B. Joo, Nucl. Phys. Proc. Suppl. {\bf 140}, 832 (2005)
[arXiv: hep-lat/0409003].

\bibitem{Pippig2013}
M. Pippig, SIAM J. Sci. Comput. {\bf 35}, C213 (2013).
  
        
\end{thebibliography}
\end{document}